\documentclass[prl, twocolumn, showpacs, preprintnumbers,amsmath,amssymb]{revtex4-1}
\usepackage{graphicx}
\usepackage{dcolumn}
\usepackage{bm}
\usepackage{subfigure}
\usepackage{multirow}
\usepackage{amsmath}
\usepackage{color}
\usepackage[font=footnotesize, justification=raggedright,singlelinecheck=false]{caption}
\newcommand{\DG}{\Delta G}
\newcommand{\Uinf}{U_{\text{inf}}}

\newcommand{\vv}[1]{\ensuremath{\mathbf{#1}}}
\newcommand{\aaa}{\ensuremath{\alpha }}
\newcommand{\bbb}{\ensuremath{\beta }}

\newcommand{\na}{n_{\aaa}}

\begin{document}

\title{Entropy stabilizes floppy crystals of mobile DNA-coated colloids}
\author{Hao Hu}
\affiliation{Chemical Engineering, School of Chemical and Biomedical Engineering, Nanyang Technological University, 62 Nanyang Drive, Singapore 637459}
\author{Pablo Sampedro Ruiz}
\affiliation{Chemical Engineering, School of Chemical and Biomedical Engineering, Nanyang Technological University, 62 Nanyang Drive, Singapore 637459}
\author{Ran Ni}
\email{r.ni@ntu.edu.sg}
\affiliation{Chemical Engineering, School of Chemical and Biomedical Engineering, Nanyang Technological University, 62 Nanyang Drive, Singapore 637459}

\begin{abstract}
Grafting linkers with open ends of complementary single-stranded DNA
makes a flexible tool to tune interactions between colloids, 
which facilitates the design of complex self-assembly structures.
Recently, it has been proposed to coat colloids with mobile DNA linkers,
which alleviates kinetic barriers without high-density grafting, 
and also allows the design of valency without patches. 
However, the self-assembly mechanism of this novel system is poorly understood.
Using a combination of theory and simulation, we obtain phase diagrams for 
the system in both two and three dimensional spaces, and find stable floppy square 
and CsCl crystals when the binding strength is strong, even in the infinite binding
strength limit. We demonstrate that these floppy phases are stabilized 
by vibrational entropy, and ``floppy" modes play an important role 
in stabilizing the floppy phases for the infinite binding strength limit.  
This special entropic effect in the self-assembly of mobile DNA-coated colloids is very different from conventional molecular self-assembly, and it offers new axis to help design novel functional materials using mobile DNA-coated colloids.
\end{abstract}

\maketitle
Nucleic acids are ubiquitous in nature because of their capability of encoding large amounts of information via canonical Watson-Crick base-paring interactions~\cite{dna}. With the help of chemical methods to make synthetic oligonucleotides of arbitrary sequences, one can use specific binding interactions between single-stranded DNA (ssDNA) chains to program selective interactions between different colloidal particles. For example, one can graft DNA linkers to the surface of colloidal particles with open ends of ssDNAs. These ssDNA tails serve as ``sticky ends'' that bind specifically to other colloids coated with ssDNA tails of complementary sequence, which offers a novel way of manipulating the self-assembly of colloidal particles~\cite{Mirkin96,Alivisatos96}. By using DNA-coated colloids (DNACCs), a number of ordered crystals~\cite{Macfarlane11,Nykypanchuk08,Casey12,Park08,Auyeung14,Wang15a,Wang15b} and self-assembled ``colloidal molecules''~\cite{Wang12} have been obtained in experiments, while the  self-assembly mechanism of DNACCs is still not well understood~\cite{Jenkins14,Wang15a,Song17}. For example, the diffusionless transformation from a floppy crystal to the other compact crystal has been observed in experimental systems while the underlying mechanism remains not fully resolved~\cite{Song17}. 

Recently, a novel system of mobile DNA-coated colloids (mDNACCs) 
was introduced that displays qualitatively new properties~\cite{Meulen13,Meulen15}. 
Compared with immobile DNA-coated colloidal systems, 
mDNACCs have a broader temperature window for self-assembly, and therefore allow the better control 
over the assembly process~\cite{Meulen13}.  
Mobility of DNA linkers also allows particles to more easily roll 
around each other and rearrange~\cite{Meulen13,Feng13}, 
without grafting of very high density. 
Moreover, unlike colloids with patches in specific locations~\cite{Wang12,Feng13}, the interaction in mDNACCs is intrinsically a many-body potential, which could be employed to control the ``valency'' of particles without patches by tuning nonspecific repulsions between the particles~\cite{Angioletti-Uberti14,Angioletti-Uberti16}.
However, despite these novel properties and potential applications, 
the principles determining the collective self-assembly of mDNACCs remain unclear. To this end, we study the equilibrium self-assembly in binary systems of mDNACCs with complementary sequences.
We construct the phase diagrams for systems in both two (2D) and three dimensional (3D) spaces.
At low pressure, we find floppy square and CsCl crystals in 2D and 3D systems, respectively, 
which are more stable than the corresponding compact hexagonal and CuAu crystals.
This behavior holds for a large range of binding strengths, even in the infinite binding strength limit. 
We demonstrate that these floppy crystals are stabilized by vibrational entropy,
and ``floppy" modes play an important role in the infinite binding strength limit. 

We consider a binary system of $N$ colloids $A$ and $B$, 
coated with mobile DNA linkers. Each linker terminates in a short ssDNA sequence, and particles of type $A$ and $B$ are coated with ssDNA linkers of complementary sequences. The parameters are chosen to be the same as for the system 
without nonspecific repulsions in Ref.~\cite{Angioletti-Uberti14}: 
the systems has equal numbers of $A$ and $B$ particles $N_A = N_B = N/2$;
each colloid is modelled as a hard sphere with diameter $\sigma = 200 \; nm$,
on which $n = 70$ double-stranded DNA (dsDNA) linkers of length $L = 20 \; nm$ 
terminating in a short ssDNA sequence are grafted.
The effective interaction energy $\beta U$ consists of an attraction part 
coming from the binding of the linkers $\beta F_{\rm att}$, 
and a repulsive part due to the excluded volume interaction $\beta F_{\rm rep}$, where $\beta = 1/k T$ with $k$ and $T$ being the Boltzmann constant and temperature of the system, respectively. 
Using a mean-field approach~\cite{Angioletti-Uberti14, Varilly12, Angioletti-Uberti13},
$\beta F_{\rm att}$ can be written as
\begin{eqnarray}
        \beta F_{\rm att} &=& \sum_{i=1}^{N} n[\ln p_i + (1-p_i)/2].
\label{eq:finite-F-att}
\end{eqnarray}
Here $p_i$ is the probability that a linker on particle $i$ is unbound, satisfying the following set of equations
\begin{eqnarray}
	p_i + \sum_j p_i p_j e^{-\beta \Delta G_{ij}} = 1 \;.
\label{eq:self-consistent-finite}
\end{eqnarray}
$\beta \Delta G_{ij}(\vv{r}_i, \vv{r}_j)$ is the free energy
for the formation of a bond between a pair of particles $i-j$, which can be written as
\begin{eqnarray}
        \beta \Delta G_{ij}(\vv{r}_i, \vv{r}_j) = \beta \Delta G_0 + \beta \Delta G_{\rm cnf}(\vv{r}_i, \vv{r}_j) \; ,
\end{eqnarray}
where $\beta \Delta G_0$ is the binding strength (hybridization free-energy) of 
two complementary ssDNAs in solution, depending on the DNA sequence and being a function of temperature and salt concentration, and $\vv{r}_{i/j}$ is the position of particle $i/j$.
The smaller value of $\beta \DG_0$ implies stronger binding strength. 
$\beta \Delta G_{\rm cnf}$ is the configurational cost for bond formation 
that can be calculated analytically for $L \ll \sigma$~\cite{Angioletti-Uberti14}.
The repulsion  originates from excluded interactions between DNA linkers 
and hard-sphere cores, and it is of general form~\cite{Angioletti-Uberti14}
\begin{eqnarray}
        \beta F_{\rm rep} &=& - \ln \left( \frac{\Omega(\{\vv{r}_i\})}{\Omega_{\rm free}} \right) \;, 
\label{eq:finite-F-rep}
\end{eqnarray}
where the partition function $\Omega(\{\vv{r}_i\})$ counts all accessible states of linkers given the
positions of the colloids, and $\Omega_{\rm free}$ is the value of $\Omega(\{\vv{r}_i\})$
when colloids are separated from each other by an infinite distance.
Then using $ \beta U = \beta F_{\rm att} + \beta F_{\rm rep}$ above, 
we perform extensive Monte Carlo simulations for systems of $N \approx 500$ mDNACCs; and by free energy calculations~\cite{FrenkelSmitBook}, we construct the phase diagrams in both 2D and 3D. In our 2D system of mDNACCs, colloids are moving in a 2D plane while DNA linkers rotate in a 3D space, which can model the self-assembly of mDNACCs at the bottom of an experimental chamber~\cite{Song17,Meulen13} or the liquid-liquid interface~\cite{Aveyard00}.

\begin{figure}
\begin{center}
\includegraphics[width=0.9\columnwidth]{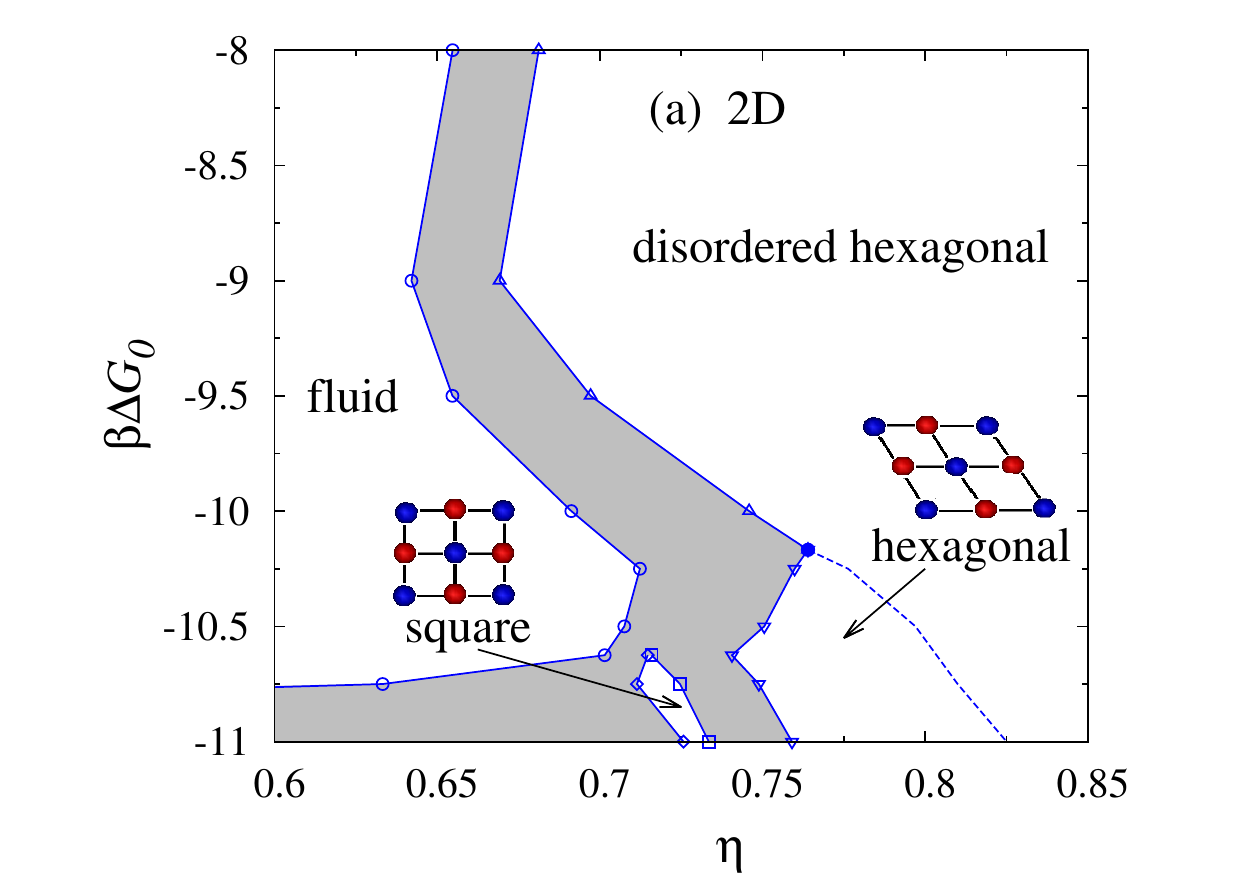} \\
\includegraphics[width=0.9\columnwidth]{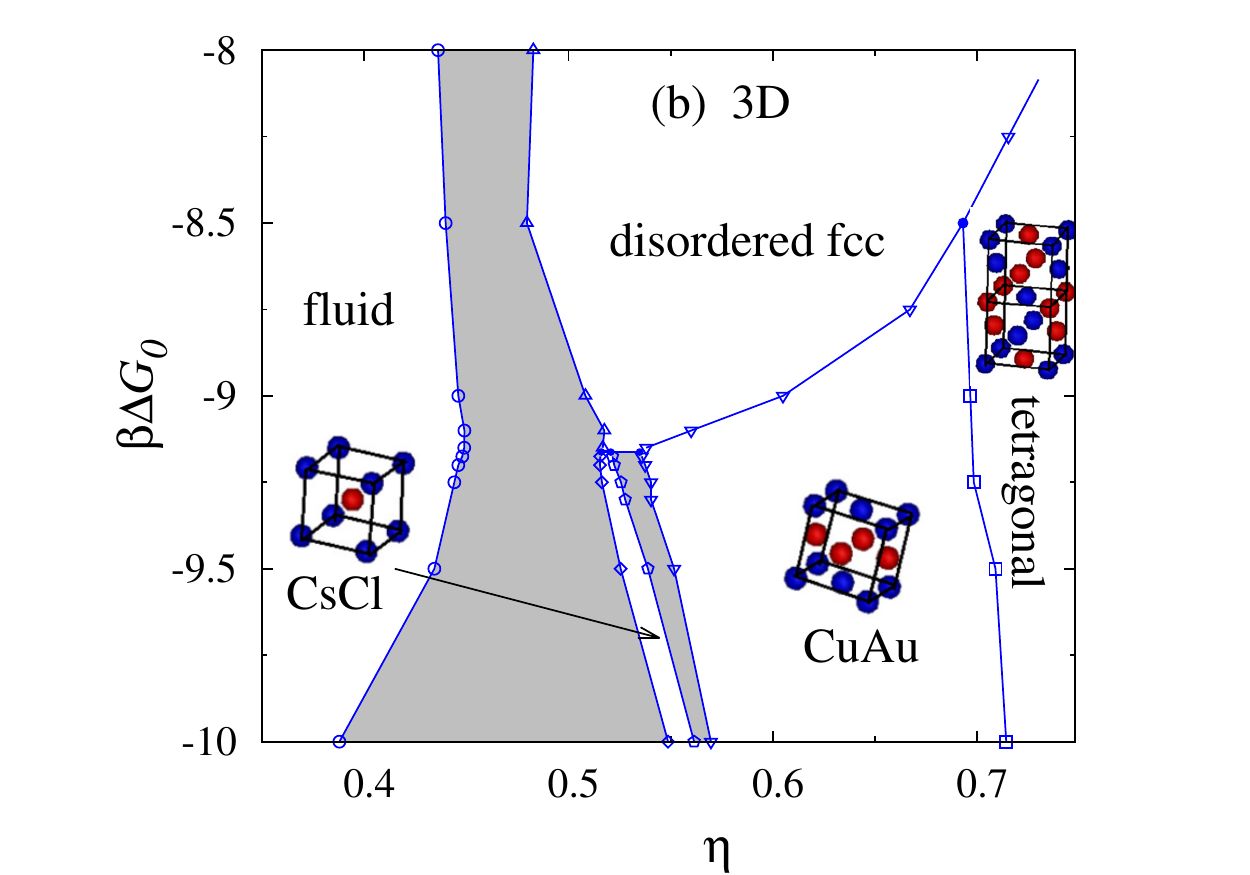}
\end{center}
\caption{(color online). Phase diagrams for mobile DNA-coated colloids in two and three dimensional space, 
	in the area/packing fraction $\eta$ - binding strength $\beta \Delta G_0$ plane. 
	Black circles mark the position of the triple points obtained by extrapolation.  
	The dashed line is an estimate of the transition from the hexagonal phase
	to the disordered hexagonal phase.
	Snapshots are shown for the ordered crystal phases. 
	Gray areas represent the coexistence regions.}
\label{fig:phase_diagram}
\end{figure}

Figure~\ref{fig:phase_diagram} shows phase diagrams 
in the area/packing fraction $\eta$ - binding strength $\beta \DG_0$ representation for mDNACC systems in 2D and 3D, respectively.
Since the numbers of complimentary linkers on both A and B colloids are the same, we do not consider crystals of asymmetric stoichiometry.
One can see that at low density, because of entropy, the systems remain in a disordered fluid phase~\cite{Bozorgui08,Martinez-Veracoechea11}. 
When increasing the density, ordered crystals form. For weak ssDNA bindings, mDNACCs crystallize into disordered crystals, i.e. disordered hexagonal crystal in 2D and face-centered cubic crystal in 3D, in which particles are located on ordered lattices but types of particles are random. When increasing the ssDNA binding strength, a few ordered crystals appear in the phase diagrams. As shown in Fig.~\ref{fig:phase_diagram}, for $\beta \Delta G_0 \lesssim -10.5$, when increasing the density of 2D mDNACC systems, a floppy ordered square crystal first crystallize from the fluid (Fig.~\ref{fig:phase_diagram}a), and similarly in 3D systems for $\beta \Delta G_0 \lesssim -9$, a floppy CsCl crystal forms at relatively low density (Fig.~\ref{fig:phase_diagram}b). When further increasing the density, the compact crystals become stable because of their high packing efficiency, i.e. hexagonal crystals in 2D, CuAu and tetragonal crystals in 3D.
The phase diagram of 3D mDNACC system is qualitatively similar to that of oppositely charged colloids~\cite{Hynninen06}.
A remarkable feature of these phase diagrams is the wide range of binding strengths for which floppy 
square and CsCl crystals are stable, given that there are many linkers on each colloidal particle and increasing the binding strength of a pair of complementary ssDNAs by $1kT$ can dramatically enhance the binding potential between colloids.

\begin{figure}
\begin{center}
\includegraphics[width=1.0\columnwidth]{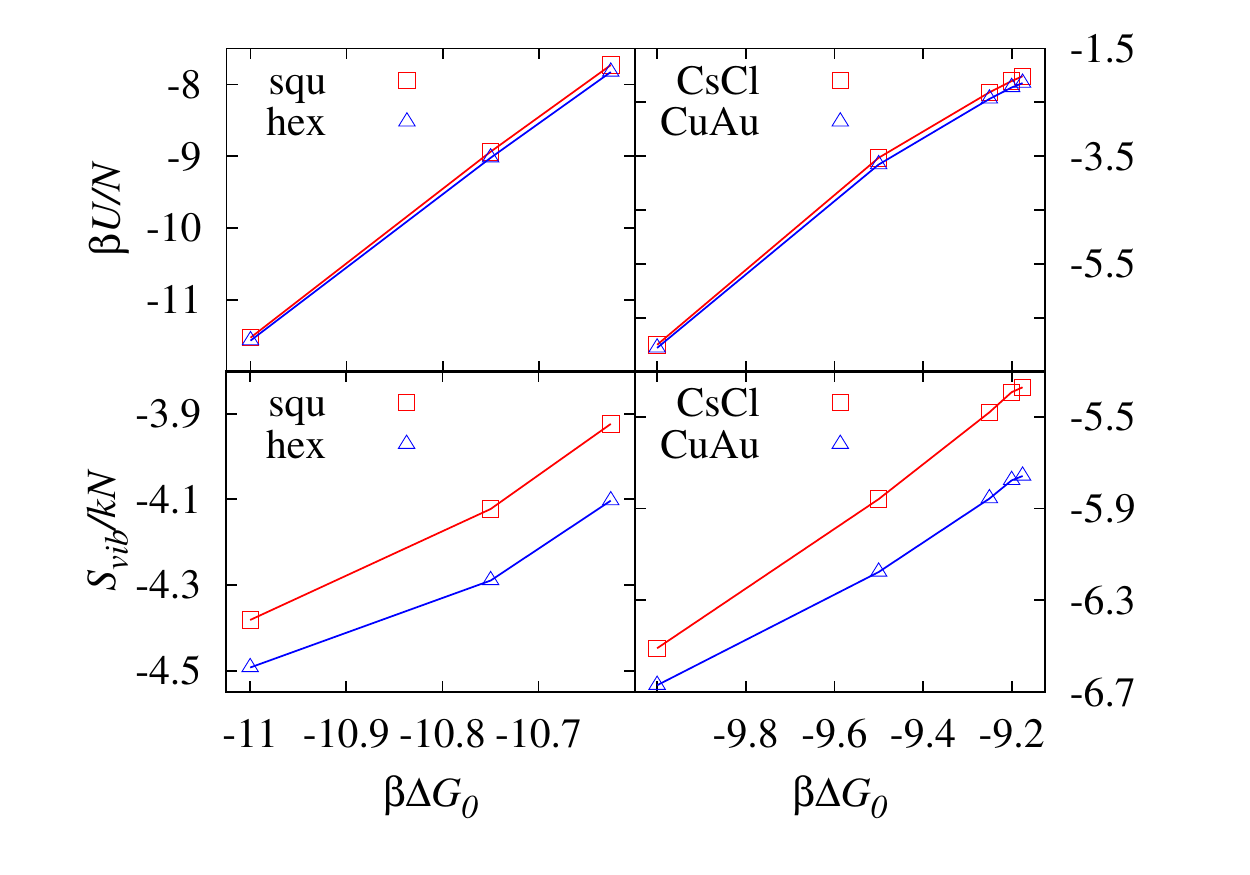} \\
\end{center}
\caption{(color online). Interaction potential $\beta U/N$ and vibrational entropy $S_{\rm vib}/kN$ 
versus the binding strength $\beta \Delta G_0$, at the square - hexagonal phase coexistence in 2D,
and CsCl - CuAu phase coexistence in 3D. }
\label{fig:U-S-finite}
\end{figure}

Stabilization of floppy crystals can be due to either enthalpy or entropy.
For a single component system of DNA-coated nanoparticles~\cite{Thaner15}, 
including the configurational entropy of the linkers in the effective potential 
(as adopted in this work), the b.c.c. crystal is believed to be favored 
over the f.c.c. crystal due to its lower enthalpy~\cite{Rogers16,Thaner15}. 
And for a binary system of colloids coated with very short DNA linkers,
it has been mentioned that the CsCl structure is favored 
over the CuAu structure by virtue of its higher vibrational entropy~\cite{Casey12}.
Despite these examples, stabilization mechanisms for floppy crystals 
are largely undistinguished for DNACCs~\cite{Martinez-Veracoechea11,Mladek12,Wang15a,Song17}.
For this case, to explore the stabilization mechanism for the floppy crystals in mDNACCs,
we compare the potential energy and vibrational entropy at coexistence. 
For simplicity, we use $S = S_{\rm vib}$ and neglect other contributions to the entropy 
since they are the same for phases being compared.
The effective potential energy $\beta U$ is calculated directly in Monte Carlo simulations.
The vibrational entropy per particle is obtained  as 
$S_{\rm vib}/kN = \beta U/N - \beta F/N$, in which the Helmholtz free energy $\beta F$
is calculated by thermodynamic integrations. 
As shown in Fig.~\ref{fig:U-S-finite},
we see that the floppy square/CsCl crystal has
higher vibrational entropy than the compact hexagonal/CuAu crystal, while the effective energy for the floppy square/CsCl crystal is even slightly higher. This suggests that the vibrational entropy stabilizes 
the observed floppy crystals in mDNACCs. 

In experiments, the hybridization free-energy 
$\beta \DG_0$ is sensitive to external conditions, such as the salt concentration
and temperature~\cite{Angioletti-Uberti16}, and tuning $\beta \DG_0$ precisely can be challenging. Therefore, when the crystallization of DNACCs is observed, the hybridization free-energy is usually very strong. To examine whether entropy stabilized floppy crystals of mDNACCs exist at conditions close to experiments, we simulate mDNACCs at $\beta \DG_0 \to -\infty$, which are essentially the very bottom of the phase diagrams in Fig.~\ref{fig:phase_diagram}.

\begin{figure*}[t]
\begin{center}
\includegraphics[width=0.8\textwidth]{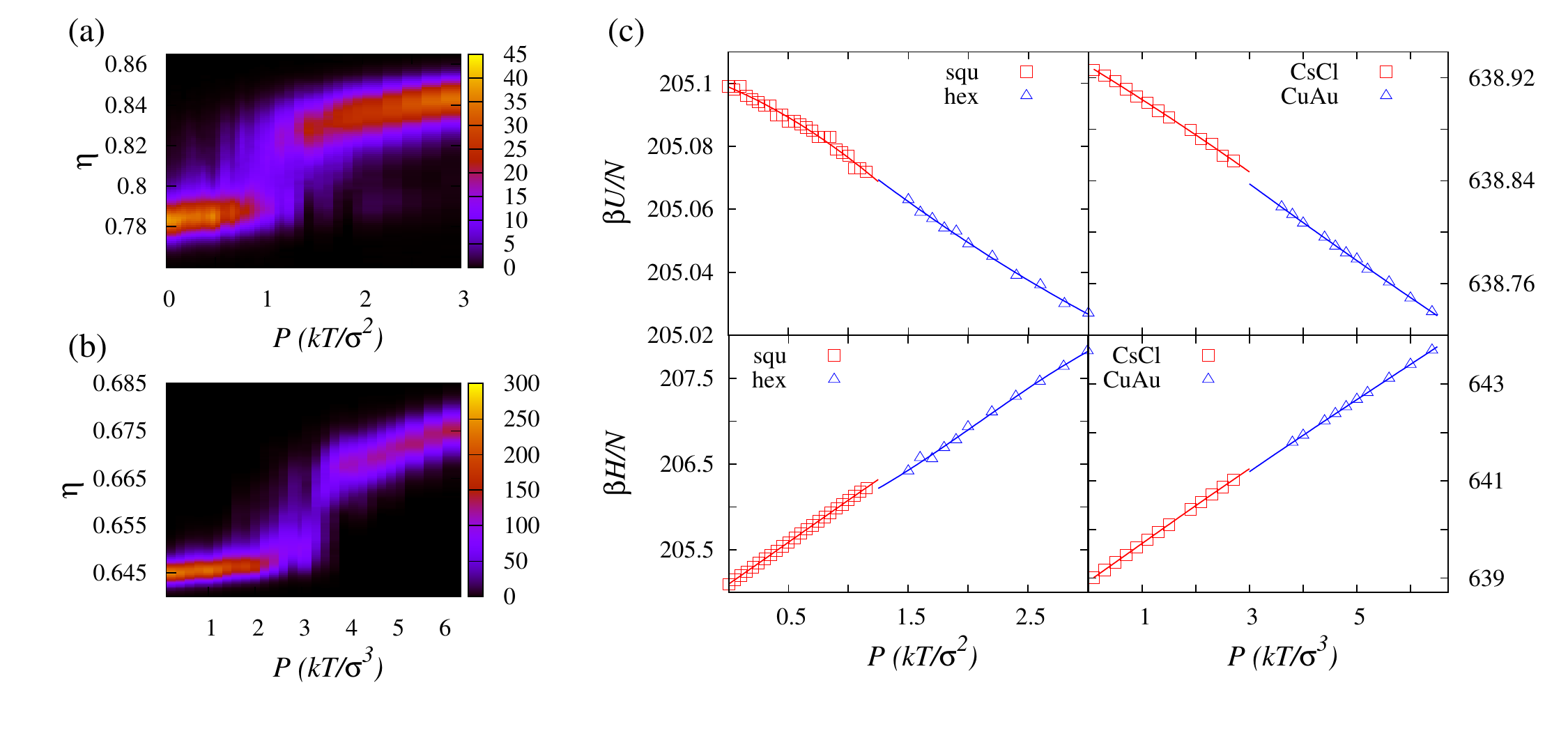} \\
\end{center}
\caption{(color online). Simulation results for mobile DNA-coated colloids at the infinite binding strength 
	limit $\beta \DG_0 \to -\infty$. 
	(a,b) Probability density distribution of area/packing fraction $\eta$ of the system versus 
	pressure $P$ in 2D (a) and 3D (b).
	(c) Interaction potential $\beta U/N$ and enthalpy $\beta H/N$ versus 
	the pressure $P$. Solid curves are obtained by data fitting.}
\label{fig:infinite}
\end{figure*}

The above method for simulating mDNACCs can only be used to simulate mDNACCs with moderate DNA hybridization free-energies and many unbound ssDNA linkers~\cite{Angioletti-Uberti14, Varilly12, Angioletti-Uberti13}.
For the limit $\beta \DG_0 \to -\infty$, Eq.~\ref{eq:finite-F-att} leads to a divergent $\beta F_{att}$. 
Therefore, we formulate a new  Monte Carlo method below to simulate mDNACCs at this limit,
with focus on the effect of entropy. 
Essentially, when $\beta \DG_0 \to -\infty$, all ssDNA linkers are bonded to complementary linkers on neighboring particles 
\begin{equation}
        \na = \sum\limits_\beta x_{\aaa\beta} \, ,
        \label{eq:na}
\end{equation}
where $\na$ is the number of linkers on particle $\aaa$; $x_{\aaa\beta}$ is the number of DNA bonds 
between particle $\aaa$ and its neighbor $\beta$; and the summation runs over all neighboring
particles of $\aaa$. 
We can then write down the partition function associated with the configuration of the DNA linkers,
and derive the effective potential energy as (details presented in the Supplemental Material~\cite{suppMat})
\begin{eqnarray}
	\beta \Uinf (\{ x_{\aaa\bbb}\}) = 
\sum_{\aaa<\bbb}
x_{\aaa\bbb} \left( 
         \ln x_{\aaa\bbb} - 1 -   \ln \Xi_{\aaa\bbb} 
 \right) - \ln Z_0 \; ,
\label{eq:ftot}
\end{eqnarray}
where $\Xi_{\aaa\bbb} = \Omega_{\aaa\bbb} / \left( \rho_0 \Omega_0^2 \right)$,
with $\rho_0$ being the standard concentration;
$\Omega_{\aaa\bbb}$ being the configuration space for two strands grafted on neighboring 
particles $\aaa$, $\bbb$ and bonded to each other; and $\Omega_0$ 
being the configuration space for an unbound linker when the particles are separated at the dilute limit; $Z_0 = \prod_{\aaa} n_{\aaa}! \exp (-\beta \DG_0 n_{\aaa}/2)$  is a constant since $n_\aaa$ is fixed.
The equilibrium linker distribution under the constraint in Eq.~\ref{eq:na} is given 
by minimizing the Lagrange function $\mathcal{L} = \beta \Uinf + \sum_{\alpha}{\lambda_{\alpha}(n_{\alpha} - \sum_{\gamma}{x_{\alpha \gamma}})}$ via 
\begin{eqnarray}
{\partial 
\over \partial x_{\aaa\bbb}} \left[\beta \Uinf (\{x_{\aaa\bbb}\}) 
+ \sum_{\aaa} \lambda_\aaa \left( n_\aaa - \sum_\gamma x_{\aaa\gamma} \right)  \right] = 0 \, .
\label{eq:sp1}
\end{eqnarray}
Substituting Eq.~\ref{eq:ftot} and $x_{\aaa\bbb} = x_{\bbb\aaa}$ into Eq.~\ref{eq:sp1}, 
we get
\begin{eqnarray}
        x_{\aaa\bbb} = e^{\lambda_{\aaa} + \lambda_{\bbb}} \Xi_{\aaa\bbb} \, ,
\label{eq:xab}
\end{eqnarray}
where coefficients $\{ \lambda_{\aaa} \}$ satisfy the constraint in Eq.~\ref{eq:na}, 
and can be solved by, e.g., self-consistent iterations with
\begin{eqnarray}
        e^{\lambda_\aaa} = {n_\aaa \over \sum_\gamma e^{\lambda_\gamma} \Xi_{\aaa\gamma}} \, , \,\, {\rm for \,\, all}\,\, \aaa \, .
\label{eq:lambda}
\end{eqnarray}
We prove that
Eq.~\ref{eq:xab} gives the global minimum of $\beta \Uinf$ under the constraint in Eq.~\ref{eq:na}~\cite{suppMat}. We then use $\beta U = \beta \Uinf + \ln Z_0$ with the constraint in Eq.~\ref{eq:na} to perform Monte Carlo simulations in the NPT ensemble 
for both 2D and 3D mDNACC systems. Results are shown in Fig.~\ref{fig:infinite} for a system of $N=100$ in 2D 
and $N=256$ in 3D. 
From the probability density distributions of packing fraction of the systems in Fig.~\ref{fig:infinite}a,b,
we see that floppy crystals do exist in the low pressure region.
Coexistence of floppy and compact phases occurs at about $P \simeq 1.0 kT/\sigma^2$ and $3.0 kT/\sigma^3$ for 2D and 3D systems, respectively.
As shown in Fig.~\ref{fig:infinite}c, near the phase coexistence, 
the effective potential per particle $\beta U/N$ for hexagonal and square crystals are very similar,
and the CuAu crystal has a slightly lower potential energy than the CsCl crystal.
Since the floppy and compact phases have equal chemical potential 
$\beta (\mu_A N_A + \mu_B N_B)/N = \beta U/N + \beta PV/N - S_{\text{vib}}/kN$ at the coexistence,
to compare the vibrational entropy of the floppy and compact phases,
we calculate the enthalpy per particle $\beta H/N = \beta U/N + \beta PV/N$.  
Figure~\ref{fig:infinite}c shows that enthalpy for the floppy phase 
is higher than that of the corresponding compact phase in both 2D and 3D near coexistence.
This implies that the floppy crystals have higher vibrational entropy near phase coexistence.
Thus vibrational entropy also stabilizes the floppy phases at the limit $\beta \DG_0 \to -\infty$.

\begin{figure}
\begin{center}
\includegraphics[width=0.8\columnwidth]{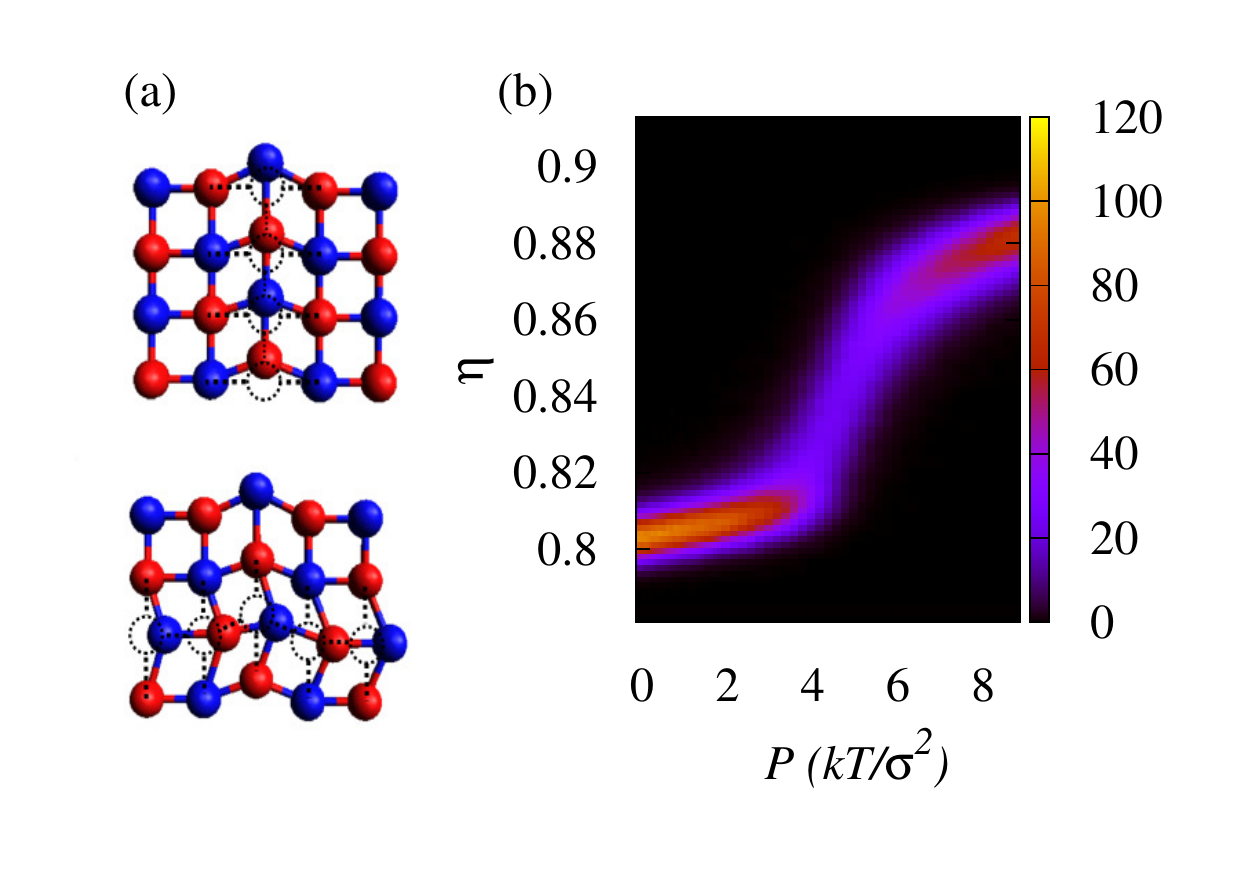} \\
\end{center}
\caption{(color online). (a) Two consecutive ``floppy" moves of the sticky-sphere model 
	from an initial square lattice. In simulations only particles in a single column 
	or row are allowed to move relatively to other particles at a time. 
	Here we show a column move in the middle (top) followed by a row move in the third row (bottom).  
	(b) Probability density distribution of area fraction $\eta$ of the system
	versus pressure $P$ for the sticky-sphere model in 2D, for which 
	distances between neighboring $A$ and $B$ particles are fixed at $d_{AB}=1.04\sigma$.}
\label{fig:EOS}
\end{figure}

A vibrational mode analysis reveals that ``floppy" modes, 
namely collective motions that do not change the interaction energy,
play an important role for the diffusionless CsCl-CuAu transition
in a system of DNACCs with short linkers~\cite{Jenkins14}.
For mDNACCs in the infinite binding limit, 
we observe that distances between neighboring $A-B$ pairs are very short ($d_{AB} \simeq 1.04 \sigma$),
and that the effective interaction energy for the floppy and compact
crystals are very similar at coexistence,
especially for the 2D system. These motivate us to explore the role of ``floppy" modes 
in the stabilization of floppy crystals.
We tried a vibrational mode analysis within the harmonic approximation, 
to evaluate the entropy~\cite{Mao13a, Mao13b} and count the floppy modes~\cite{Jenkins14}.
We calculated approximately the dynamical matrix by measuring displacement correlations 
between particles~\cite{Chen10}. However, for system sizes we can currently simulate, 
we observe large volume fluctuations which invalidate the harmonic approximation.
Instead, we approximate the system by a sticky-sphere model: 
every colloid is bonded with a fixed number of colloids of the opposite type 
and distances between all $A-B$ pairs are fixed. This approximation is exact for mDNACC systems with $\beta \DG_0 \to -\infty$ and linker length $L \to 0$, for which the effective potential $\beta \Uinf$ is a constant 
and the system allows only ``floppy" moves.
We conducted NPT simulations for the model with $N=100$ in 2D.
To illustrate how the model works,
in Fig.~\ref{fig:EOS}a we show two consecutive moves from an initial square lattice.
The probability density distribution of area fraction of the system is shown in Fig.~\ref{fig:EOS}b
for different pressure. We see that, for pressure below $P \simeq 4 kT/\sigma^2$,
``floppy" moves favor the low density square crystal over the dense hexagonal crystal.
This result supports that vibrational entropy associated with ``floppy" modes 
play an important role in the stabilization of observed floppy phases 
in the infinite binding limit.
This is related to the fact that mechanically floppy networks can become rigid (stable) 
when thermal effects are present~\cite{Dennison13}.

In conclusion, we have studied the self-assembly of a binary system of colloids coated with complementary mobile DNA linkers. 
We construct the phase diagrams for both 2D and 3D mDNACC systems,
in which we observe stable floppy square/CsCl crystals at strong enough ssDNA binding strength.
We also derive an effective potential for the system in the infinite ssDNA binding strength limit, and formulate a Monte Carlo method to simulate the effect of entropy for the system in this limit.
Our results show that even for the infinitely strong ssDNA binding limit, floppy crystals are still more stable than compact ones at low pressure because of their higher vibrational entropy.
This suggests that the strong ssDNA binding limit of mDNACC systems is different from the conventional atomistic or molecular systems, in which at strong interaction limit, i.e. zero temperature limit, the effect of entropy vanishes. Although we simulated the system of colloids coated with mobile DNA linkers, the observed special effect of vibrational entropy can also explain the physics of forming floppy CsCl crystal of colloids coated with long flexible DNA linkers at experimental conditions, in which the binding sites of DNA linkers can cover the whole surface of DNA corona~\cite{Mladek12}.
This opens up new possibilities for designing mDNACC systems to utilize the effect of entropy for fabrication of novel functional colloidal materials.
For example, in asymmetric binary mixtures of mDNACCs, more open structures are expected, especially when nonspecific repulsions are introduced ~\cite{Angioletti-Uberti14}, of which the self-assembly mechanism can be very different from that of conventional patchy particle systems~\cite{Chen11,Mao13a,Mao13b}.
Moreover, in experimental systems of mobile DNA coated liposomes, the combination of vibrational entropy and deformability are expected leading to the formation of more interesting structures~\cite{Shimobayashi15,mognetti1,mognetti2,mognetti3}.
Additionally, 
the method developed for the infinite binding strength limit could also be modified to simulate other network systems, e.g. vitrimers~\cite{Smallenbury13,Romano15}, in which the relaxation is not driven by energy change but a result of entropy maximization.

\begin{acknowledgments}
The authors acknowledge Dr. Qunli Lei for helpful discussions. This work is supported by Nanyang Technological University Start-Up Grant (NTU-SUG: M4081781.120), the Academic Research Fund Tier 1 from Singapore Ministry of Education (M4011616.120 and M4011873.120), and the  Advanced Manufacturing and Engineering Young Individual Research Grant (M4070267.120) by the Science \& Engineering Research Council of Agency for Science, Technology and Research Singapore.
We are grateful to the National Supercomputing Centre (NSCC) of Singapore for supporting the numerical calculations.
\end{acknowledgments}

\bibliographystyle{h-physrev}
\bibliography{ref}
\clearpage
\newcommand{\beginsupplement}{%
        \setcounter{table}{0}
        \renewcommand{\thetable}{S\arabic{table}}%
        \setcounter{figure}{0}
        \renewcommand{\thefigure}{S\arabic{figure}}%
        \setcounter{equation}{0}
        \renewcommand{\theequation}{S\arabic{equation}}%
     }
\onecolumngrid
\beginsupplement

\section{Supplementary Materials}
As explained by Angioletti-Uberti et al. [in the Supplemental Material of PRL {\bf 113}, 128303 (2014)],
for mobile DNA-coated colloids at fixed positions, the partition function accounting
for the binding of DNA is given by
\begin{eqnarray}
        Z_{\rm bind} &=&\sum_{\{x_{\aaa\bbb}\}}  W(\{x_{\aaa\bbb}\}) \prod\limits_{\aaa<\bbb} \Theta_{\aaa\beta}^{x_{\aaa\bbb}} \, ,
\nonumber \\
W(\{x_{\aaa\bbb}\}) &=& \prod_\alpha {\na ! \over (\na-\sum\limits_\beta x_{\aaa\beta})!} \prod\limits_{\aaa<\beta} {1\over x_{\aaa\beta}!} \, ,
\label{Seq:z1}
\end{eqnarray}
where $W (\{ x_{\alpha \beta} \})$ counts all the possible combinations of DNA-DNA hybridisation which lead to $x_{\alpha\beta}$ bonds between particle $\alpha$ and $\beta$, and $\na$ is the number of DNA linkers on particle $\alpha$.
In the infinite binding strength limit $\beta \DG_0 \to -\infty$, all linkers are bound
\begin{equation}
        \na = \sum\limits_\beta x_{\aaa\beta} \, ,
        \label{Seq:na}
\end{equation}
where the summation is over nearest neighbors which can bind to particle $\alpha$.
The bond strength $\Theta$ is given by
\begin{align}
        \Theta_{\aaa\bbb}\left( \vv{R}_A,\vv{R}_B \right) &= <\exp\left( -\beta \DG_{\aaa\bbb} \right)>_{\mid \vv{R}_A,\vv{R}_B} \nonumber \\
            &= { \int_{S_A,S_B} \exp\left[-\beta\DG_{\aaa\bbb}\right]d\vv{r}_\aaa d\vv{r}_\bbb \over S_A S_B } \nonumber \\
            &= \exp\left[-\beta\DG_0\right] { \int_{S_A,S_B} \exp\left[-\beta\DG_{\rm cnf}(\vv{r}_\aaa, \vv{r}_\bbb)\right]d\vv{r}_\aaa d\vv{r}_\bbb \over S_A S_B } \nonumber \\
     &= \Theta_0 \Theta^{*}_{\aaa\bbb} \, ,
        \label{Seq:bond-average}
\end{align}
where the average is taken with the center of colloid $A(B)$ fixed at $\vv{R}_{A(B)}$;
$S_{A(B)}$ is the surface area of colloid;
and $\Theta_0 = \exp\left[-\beta\DG_0\right]$; $\vv{r}_{\alpha(\beta)}$ is the grafting position of linkers on particle $\alpha(\beta)$.
$\beta \Delta G_{\aaa\bbb} = \beta \Delta G_0 + \beta \Delta G_{\rm cnf}(\vv{r}_\aaa, \vv{r}_\bbb)$ gives the free energy for the formation of a bond $\aaa-\bbb$,
with $\beta \Delta G_{\rm cnf}(\vv{r}_\aaa, \vv{r}_\bbb)$
being the configurational cost associated with the bond formation.
Substituting Eqs.~\ref{Seq:na},~\ref{Seq:bond-average} into Eq.~\ref{Seq:z1}, we get
\begin{eqnarray}
        Z_{\rm bind} &=\sum\limits_{\{x_{\aaa\bbb}\}}  \prod\limits_\alpha {\na !} \prod\limits_{\aaa<\beta} {1\over x_{\aaa\beta}!} \prod\limits_{\aaa<\bbb} \left(\Theta_0 \Theta_{\aaa\beta}^{*}\right)^{x_{\aaa\bbb}} \nonumber \\
         &=\sum\limits_{\{x_{\aaa\bbb}\}}  \prod\limits_\alpha {\na !} \prod\limits_{\aaa<\beta} \Theta_0^{x_{\aaa\bbb}} \prod\limits_{\aaa<\bbb} {1\over x_{\aaa\beta}!} \Theta_{\aaa\beta}^{* x_{\aaa\bbb}} \, .
\label{Seq:z2}
\end{eqnarray}
Since the number of linkers $\na$ is fixed, the prefactor
$Z_0 \equiv \prod\limits_\alpha {\na !} \prod\limits_{\aaa<\beta} \Theta_0^{x_{\aaa\bbb}} = \prod_{\aaa} n_{\aaa}! \exp (-\beta \DG_0 n_{\aaa}/2)$
contributes an infinite constant.
Thus the partition function can be written as
\begin{eqnarray}
        Z_{\rm bind} &=& Z_0 \sum\limits_{\{x_{\aaa\bbb}\}} \prod\limits_{\aaa<\bbb} {1\over x_{\aaa\beta}!} \Theta_{\aaa\beta}^{* x_{\aaa\bbb}} \, .
\label{Seq:z3}
\end{eqnarray}
Using Stirling's approximation, we can express the partition function as
\begin{eqnarray}
        Z_{\rm bind} &=& \sum_{\{ x_{\aaa\bbb}\}} e^{-\beta U_{\rm bind} (\{ x_{\aaa\bbb}\})} \; ,
\label{Seq:ffun}
\\
\beta U_{\rm bind} (\{ x_{\aaa\bbb}\}) &=& 
\sum_{\aaa<\bbb} x_{\aaa\bbb}
\Big(
         \ln x_{\aaa\bbb} - 1 -  \ln \Theta_{\aaa\bbb}^{*}
\Big) - \ln Z_0 \; .
\nonumber
\end{eqnarray}
The above binding energy is accurate when the number of linkers is large,
since Stirling's approximation is valid only for large values of $x_{\aaa\bbb}$.

The binding energy tells the free-energy difference between the binding state
and the nonbinding. The free energy of the later is a purely
repulsive energy of the form
\begin{eqnarray}
        \beta U_{\rm rep} = - \sum_{\aaa}  
         n_\aaa \ln \frac{\Omega_\aaa}{\Omega_0}  \, ,
\end{eqnarray}
where $\Omega_\aaa$ is the phase space allowed for an unbound linker
on particle $\aaa$, and $\Omega_0$ is the phase space
allowed for the same linker when particle $\aaa$ is separated from
other particles by an infinite distance.

The total free energy is given by
\begin{eqnarray}
        \beta \Uinf = \beta U_{\rm bind}  + \beta U_{\rm rep} \, .  
\end{eqnarray}
Since $\Theta_{\aaa\bbb}^{*} = \Omega_{\aaa\bbb} / \left( \rho_0 \Omega_\aaa \Omega_\bbb \right)$ where $\Omega_{\aaa\bbb}$ is the phase space allowed for two mobile linkers
grated on particles $\aaa$ and $\bbb$ when they are bound to each other
[see the Supplemental Material of Angioletti-Uberti, PRL {\bf 113}, 128303 (2014)],
we obtain
\begin{eqnarray}
\beta \Uinf (\{ x_{\aaa\bbb}\}) = 
\sum_{\aaa<\bbb}
 x_{\aaa\bbb} \Big(
         \ln x_{\aaa\bbb} -1 -  \ln \Xi_{\aaa\bbb}
\Big) - \ln Z_0 \; ,
\label{Seq:ftot}
\end{eqnarray}
where $\Xi_{\aaa\bbb} = \Omega_{\aaa\bbb} / \left( \rho_0 \Omega_0^2 \right)$.

It follows that the Hessian of $\beta \Uinf$ for an arbitrary configuration is
\begin{equation}
        \frac{\partial^2 \beta \Uinf}{\partial x_{\aaa\bbb} \partial x_{\aaa'\bbb'}} =
        \begin{cases}
        1/x_{\aaa\bbb}\;, & \text{if} \ \aaa=\aaa' \ \text{and} \ \bbb = \bbb' \;, \\
        0\;,            & \text{otherwise}.     
        \end{cases}
\label{Seq:Hessian}
\end{equation}
Thus the Hessian matrix $\left [ \frac{\partial^2 \beta \Uinf}{\partial x_{\aaa\bbb} \partial x_{\aaa'\bbb'}} \right ]$  is always positive definite, and $\beta \Uinf$ is a convex function. Moreover, as the constraints of Eq.~\ref{Seq:na} are linear, to minimize $\beta \Uinf$ with constraints in Eq.~\ref{Seq:na} is essentially a convex optimization problem, of which the only one local minimum is the global minimum [R. Tyrrell Rockafellar, Lagrange Multipliers and Optimality, SIAM Review {\bf 35} (2), 183 (1993)]. This implies that the solution to Eq. 7 in the main text is the global minimum of $\beta \Uinf$ subject to the linear constraints in Eq.~\ref{Seq:na}.

\end{document}